\documentclass[12pt,epsf]{article}

\usepackage{graphicx}
\usepackage{indentfirst}
\usepackage{cite}
\usepackage{color}
\usepackage{amsmath}
\begin{document}

\title{\bf{Note on the speed of GW150914 in gravity's rainbow}}

\date{}
\maketitle

\begin{center}
\author{Bogeun Gwak}$^a$\footnote{rasenis@sogang.ac.kr},
\author{Wontae Kim}$^b$\footnote{wtkim@sogang.ac.kr},
\author{Bum-Hoon Lee}$^{a,b,c,d}$\footnote{bhl@sogang.ac.kr}
\vskip 0.25in
$^{a}$\it{Center for Quantum Spacetime, Sogang University, Seoul 04107, Republic of Korea}\\
$^{b}$\it{Department of Physics, Sogang University, Seoul 04107, Republic of Korea}\\
$^{c}$\it{Asia Pacific Center for Theoretical Physics, Pohang 37673, Republic of Korea}\\
$^{d}$\it{Department of Physics, Postech, Pohang 37673, Republic of Korea}\end{center}
\vskip 0.6in

{\abstract
{
Without breaking Lorentz invariance, we investigate the speed of graviton
in event GW150914 by
using the modified dispersion relation from gravity's rainbow.
The proper range of the parameter in the modified dispersion relation
is determined by taking into account the gap between the speed of the graviton and that obtained from event GW150914.
}

\thispagestyle{empty}
\newpage
\setcounter{page}{1}
\section{Introduction}

Gravitational waves have been theoretically predicted as a solution to the linearized weak-field equation in general relativity. The wave, moving at the speed of light, could be generated from the quadrupole moment of a massive source \cite{Einstein:2016a,Einstein:2016b}; however, the amplitude of the wave was expected to be extremely small at that time, so the detectors should be sensitive enough to observe the wave. Subsequently,
it was found that long-baseline broadband laser interferometers could
 satisfy the sensitivity to detect gravitational waves, which led to the construction of the Laser Interferometer Gravitational-Wave Observatory (LIGO)\cite{Abramovici:1992ah}. Eventually, the Advanced LIGO with improved sensitivity and performance succeeded in detecting the signal of event GW150914\cite{Abbott:2016blz}. Precisely, the signal from a luminosity distance of $410^{+160}_{-180}\text{Mpc}$ has a frequency range from $35$ to $250\,\text{Hz}$, and the amplitude reaches the maximum at $150\,\text{Hz}$. The initial black hole system consists of two black holes with masses of $36^{+5}_{-4}M_\odot$ and $29^{+4}_{-4}M_\odot$. After the ringdown of a single black hole, the mass of the final black hole becomes $62^{+4}_{-4}M_\odot$, such that in the process of the binary merger, the released energy of the gravitational wave amounts to $3.0^{+0.5}_{-0.5}M_\odot c^2$. From event GW150914, the LIGO Collaboration itself~\cite{Abbott:2016blz} has suggested that the upper bound on the mass of the graviton is $m_g < 1.2 \times 10^{22} eV/c^2$, so it implies that
 the observations of the gravitational wave might be used as a constraint to quantum physics.

In the quantum regime,
assuming there exists a minimal length at the Planck scale,
special relativity could be extended to the doubly special relativity \cite{AmelinoCamelia:2000ge,AmelinoCamelia:2000mn,AmelinoCamelia:2003uc,AmelinoCamelia:2003ex,Magueijo:2001cr,Magueijo:2002am},
in which both the Planck length and the speed of light are required to be
invariant under any inertial frames.
To maintain the double invariant constants, a deformed Lorentz symmetry could
be realized by making use of a nonlinear Lorentz transformation, which yields the modified dispersion relation (MDR). On the other hand, the MDR has also been formulated on the backgrounds of gravity by using gravity's rainbow\cite{Magueijo:2002xx}, where the spacetime is probed by the test particles so that
its geometry is characterized by the energy of the test particles. Along the line of gravity's rainbow, there have been extensive studies on black holes and cosmology\cite{Galan:2005ju, Hackett:2005mb, Aloisio:2005qt, Ling:2005bp, Galan:2006by, Ling:2006az, Amelino-Camelia:2013wha, Barrow:2013gia, Ling:2008sy, Garattini:2011hy, Girelli:2006fw, Garattini:2011fs, Liu:2007fk, Peng:2007nj, Li:2008gs, Ali:2014xqa,Awad:2013nxa}.

The possibility of the Lorentz violation might be constrained by the bound on the mass of the graviton using the speed of the propagation for the graviton\cite{Ellis:2016rrr}.
The speed of the propagation for the graviton was
due to the bound on the mass. The bound on the propagation for the graviton could be originated
from the effect of the MDR having the Lorentz violation. Using event GW150914, various types of MDRs for the Lorentz violation
have been tested by using the bound on the propagation,
and then the quantum-mechanical constraints were specified by the MDR \cite{Arzano:2016twc}.
In this work, instead of breaking of Lorentz invariance, we will
investigate the speed of the graviton in event GW150914
by using gravity's rainbow\cite{Magueijo:2002xx}.
Given by the bound on the mass of the graviton from event GW150914,
the speed of the graviton can be applied to specify the range of the parameter of the MDR\cite{AmelinoCamelia:1997gz,AmelinoCamelia:2008qg,AmelinoCamelia:1996pj} in the regime of
gravity's rainbow. We will show that the effect of gravity's rainbow is very
small in low energy event of GW150914.
The paper is organized as follows. In section~\ref{sec2a}, we will introduce gravity's rainbow. In section~\ref{sec3b}, the effect of gravity's rainbow will be specified by the difference between the speeds of the graviton and light. In section~\ref{sec5}, we briefly summarize our results.

\section{Review of Gravity's Rainbow}\label{sec2a}

Doubly special relativity is the class of theories in which the Planck length is also suggested invariant under any inertial frames, as well as the speed of light. Doubly special relativity is described under the modified set of principles in special relativity\cite{Magueijo:2001cr}. The principles are:\\
(1) The relativity of inertial frames.\\
(2) In the limit of low energy relative to the Planck energy, the speed of massless particle becomes a universal constant, $c$ for all inertial observers.\\
(3) The Planck energy is also a universal constant, $\omega_p$, for all inertial observers.\\

Based on these principles, gravity's rainbow is suggested to generalize doubly special relativity to the frame of general relativity. In gravity's rainbow, the spacetime geometry probed by a test particle is deformed along the particle energy $\omega$. In the deformed spacetime, the dispersion relation of the particle having mass $m$ and momentum $p$ is also modified to the MDR\cite{Magueijo:2002xx}
\begin{eqnarray}
\label{MDR03}
\omega^2g_1(\omega/\omega_p)^2-(pc)^2 g_2(\omega/\omega_p)^2=(mc^2)^2\,,
\end{eqnarray}
where $\omega_p$ is the Planck energy. 
There are two principles to be emphasized in Gravity's rainbow\cite{Magueijo:2002xx}. These are\\
(1) Modified equivalence principle: For a large radius of curvature $R$ relative to the inverse of Planck energy
\begin{eqnarray}
R\gg \frac{1}{\omega_p}\,,
\end{eqnarray}
the laws of physics are observed to be the same as those of the modified special relativity to first order
in $\frac{1}{R}$ for the freely falling observer who measures a particle with energy $\omega$, so that
\begin{eqnarray}
\frac{1}{R}\ll \omega \ll \omega_p\,.
\end{eqnarray}
Freely falling observers to first order in 1/R can be inertial observers in rainbow flat spacetime. They describe the spacetime in terms of the family of energy dependent
orthonormal frames that
\begin{eqnarray}
e_0=\frac{1}{g_1(\omega / \omega_p)}\,,\quad e_1=\frac{1}{g_2(\omega/\omega_p)}\,,
\end{eqnarray}
in which the metric is given by
\begin{eqnarray}
g(\omega)=\eta^{ab} e_a \otimes e_b\,.
\end{eqnarray}
(2) Correspondence principle: In the limit of low-energy, $\omega/\omega_p \ll 1$, the theory becomes ordinary general relativity, so that
\begin{eqnarray}
\lim_{\omega/\omega_p\rightarrow 0} g_{ab}(E)=g_{ab}\,.
\end{eqnarray}
The MDR is determined by the rainbow functions $g_1(\omega/\omega_p)$ and $g_2(\omega/\omega_p)$, 
where they depend upon specific models. Without the particle, the MDR reduces to 
that of special relativity, so the rainbow functions should satisfy $\lim_{\omega \rightarrow 0}g_1(\omega/\omega_p)=1$ and $\lim_{\omega \rightarrow 0}g_2(\omega/\omega_p)=1$. 

We choose phenomenologically motivated rainbow functions\cite{AmelinoCamelia:1997gz,AmelinoCamelia:2008qg,AmelinoCamelia:1996pj} as
\begin{eqnarray}
\label{eta01}
g_1(\omega/\omega_p)=1\,,\quad g_2(\omega/\omega_p)=\sqrt{1-\eta \left(\frac{\omega}{\omega_p}\right)^n}\,,
\end{eqnarray}
where the free parameter $\eta$ is positive, and the exponent $n$ is positive integer.

\section{The Speed of Graviton in Event GW150914}\label{sec3b}

We will specify the parameter $\eta$ in the MDR of gravity's rainbow. 
First, the Lorentz invariant dispersion relation for the massive graviton is
\begin{eqnarray}
\omega^2=(pc)^2+(m_g c^2)^2\,,
\end{eqnarray}
where the mass of the graviton is $m_g$. The speed of the graviton is given by the group velocity of the wave front\cite{Arzano:2016twc}
\begin{eqnarray}\label{gvelo1}
v_g=\frac{d\omega}{d p}\approx c\left(1-\frac{1}{2}\left(\frac{m_g c^2}{\omega}\right)^2\right)\,,
\end{eqnarray}
where the speed of the graviton is slower than that of the light. The energy of the graviton will be assumed to be $250\text{Hz}$, which is the maximum energy in event GW150914. The difference between the speed of the graviton and light is constrained by event GW150914
\begin{align}
\Delta v_g= c-v_g=\left(\frac{m_g^2c^4}{2\omega^2}\right)c<2.0\times 10^{-12}\,\text{m/s},
\end{align}
where the graviton mass is obtained as $m_g < 1.2 \times 10^{22} eV/c^2$\cite{Abbott:2016blz}.
On the other hand, the MDR of the massless particle from Eqs.~(\ref{MDR03}) and (\ref{eta01}) is given as
\begin{eqnarray}
&\omega^2 -(pc)^2 \left(1-\eta \left(\frac{\omega}{\omega_p}\right)^n\right)=0.
\end{eqnarray}
The speed of the massless graviton is obtained from the group velocity as
\begin{align}
v=\frac{d \omega}{d p}\approx c -\eta \left(\frac{\omega}{\omega_p}\right)^n c-\frac{\eta n c}{2} \left(\frac{\omega}{\omega_p}\right)^{n}\,.
\end{align}
The speed of the massless graviton is slower than that of light. From the upper bound on $\Delta v_g$, the inequality can be set to
\begin{eqnarray}\label{eq:eta12}
\Delta v=\eta\left(1+\frac{n}{2}\right)\left(\frac{\omega}{\omega_p}\right)^nc=\Delta v_g<2.0\times 10^{-12}\,\text{m/s}\,,
\end{eqnarray}
where the deviation depends on the particle energy.
Choosing $n=2$ in Eq.~(\ref{eq:eta12}), we can fix the parameter $\eta$ at the frequency $250\text{Hz}$ as
\begin{eqnarray}
0<\eta\leq 4.6\times 10^{59}\quad.
\end{eqnarray}
The upper limit of the rainbow effects is effectively given as
\begin{eqnarray}\label{rainbow23}
\eta \left(\frac{\omega}{\omega_p}\right)^n\leq 3.3\times 10^{-21}\,,
\end{eqnarray}
where the rainbow effects become very small. 

\section{Conclusion}\label{sec5}
From the event GW150914, in order to explain the reason why the 
speed of the gravitational wave is smaller than that of the light,
we determined the parameter $\eta$ in gravity's rainbow
without introducing the graviton mass,
where the parameter $\eta$ was specified as smaller than $4.6\times 10^{59}$. 
However, this is not a large number, because the rainbow effect is effectively estimated to be lower than $3.3\times 10^{-21}$.

{\bf Acknowledgments}

{\small
BG was supported by Basic Science Research Program through the National Research Foundation of Korea (NRF) funded by the Ministry of Science, ICT {\&} Future Planning (NRF-2015R1C1A1A02037523). WK was supported by the National Research Foundation of Korea (NRF) grant funded by the Korea government (MSIP) (2014R1A2A1A11049571). BHL was supported by the National Research Foundation of Korea (NRF) grant funded by the Korea government (MSIP, 2014R1A2A1A01002306).


\begin{thebibliography}{99}

\bibitem{Einstein:2016a}
A. Einstein, Sitzungsber. K. Preuss. Akad. Wiss. {\bf 1}, 688
(1916).

\bibitem{Einstein:2016b}
A. Einstein, Sitzungsber. K. Preuss. Akad. Wiss. {\bf 1}, 154
(1918).

\bibitem{Abramovici:1992ah}
  A.~Abramovici {\it et al.},
  Science {\bf 256}, 325 (1992).

\bibitem{Abbott:2016blz}
  B.~P.~Abbott {\it et al.} [LIGO Scientific and Virgo Collaborations],
  Phys.\ Rev.\ Lett.\  {\bf 116}, no. 6, 061102 (2016).

\bibitem{AmelinoCamelia:2000ge}
  G.~Amelino-Camelia,
  Phys.\ Lett.\ B {\bf 510}, 255 (2001).


\bibitem{AmelinoCamelia:2000mn}
  G.~Amelino-Camelia,
  Int.\ J.\ Mod.\ Phys.\ D {\bf 11}, 35 (2002).

\bibitem{Magueijo:2001cr}
  J.~Magueijo and L.~Smolin,
  Phys.\ Rev.\ Lett.\  {\bf 88}, 190403 (2002).

\bibitem{Magueijo:2002am}
  J.~Magueijo and L.~Smolin,
  Phys.\ Rev.\ D {\bf 67}, 044017 (2003).

\bibitem{AmelinoCamelia:2003uc}
  G.~Amelino-Camelia,
  gr-qc/0309054.

\bibitem{AmelinoCamelia:2003ex}
  G.~Amelino-Camelia, J.~Kowalski-Glikman, G.~Mandanici and A.~Procaccini,
  Int.\ J.\ Mod.\ Phys.\ A {\bf 20}, 6007 (2005).







\bibitem{Magueijo:2002xx}
  J.~Magueijo and L.~Smolin,
  Class.\ Quant.\ Grav.\  {\bf 21}, 1725 (2004).


\bibitem{Galan:2005ju}
  P.~Galan and G.~A.~Mena Marugan,
  Phys.\ Rev.\ D {\bf 72}, 044019 (2005).

\bibitem{Hackett:2005mb}
  J.~Hackett,
  Class.\ Quant.\ Grav.\  {\bf 23}, 3833 (2006).

\bibitem{Aloisio:2005qt}
  R.~Aloisio, A.~Galante, A.~Grillo, S.~Liberati, E.~Luzio and F.~Mendez,
  Phys.\ Rev.\ D {\bf 73}, 045020 (2006).

\bibitem{Ling:2006az}
  Y.~Ling,
  JCAP {\bf 0708}, 017 (2007).

\bibitem{Girelli:2006fw}
  F.~Girelli, S.~Liberati and L.~Sindoni,
  Phys.\ Rev.\ D {\bf 75}, 064015 (2007).

\bibitem{Peng:2007nj}
  J.~-J.~Peng and S.~-Q.~Wu,
  Gen.\ Rel.\ Grav.\  {\bf 40}, 2619 (2008).

\bibitem{Ling:2008sy}
  Y.~Ling and Q.~Wu,
  Phys.\ Lett.\ B {\bf 687}, 103 (2010).



\bibitem{Garattini:2011hy}
  R.~Garattini and G.~Mandanici,
  Phys.\ Rev.\ D {\bf 85}, 023507 (2012).

\bibitem{Garattini:2011fs}
  R.~Garattini and F.~S.~N.~Lobo,
  Phys.\ Rev.\ D {\bf 85}, 024043 (2012).

\bibitem{Amelino-Camelia:2013wha}
  G.~Amelino-Camelia, M.~Arzano, G.~Gubitosi and J.~Magueijo,
  Phys.\ Rev.\ D {\bf 88}, no. 4, 041303 (2013).

\bibitem{Barrow:2013gia}
  J.~D.~Barrow and J.~Magueijo,
  arXiv:1310.2072 [astro-ph.CO].


\bibitem{Ling:2005bp}
  Y.~Ling, X.~Li and H.~-b.~Zhang,
  Mod.\ Phys.\ Lett.\ A {\bf 22}, 2749 (2007).

\bibitem{Galan:2006by}
  P.~Galan and G.~A.~Mena Marugan,
  Phys.\ Rev.\ D {\bf 74}, 044035 (2006).

\bibitem{Liu:2007fk}
  C.~-Z.~Liu and J.~-Y.~Zhu,
  Gen.\ Rel.\ Grav.\  {\bf 40}, 1899 (2008).

\bibitem{Li:2008gs}
  H.~Li, Y.~Ling and X.~Han,
  Class.\ Quant.\ Grav.\  {\bf 26}, 065004 (2009).

\bibitem{Ali:2014xqa}
  A.~F.~Ali,
Phys.\ Rev.\ D {\bf 89}, 104040 (2014).  

\bibitem{Awad:2013nxa}
  A.~Awad, A.~F.~Ali and B.~Majumder,
  JCAP {\bf 1310}, 052 (2013).


\bibitem{Ellis:2016rrr}
  J.~Ellis, N.~E.~Mavromatos and D.~V.~Nanopoulos,
  arXiv:1602.04764 [gr-qc].

\bibitem{Arzano:2016twc}
  M.~Arzano and G.~Calcagni,
  Phys.\ Rev.\ D {\bf 93}, no. 12, 124065 (2016).


\bibitem{AmelinoCamelia:2008qg}
  G.~Amelino-Camelia,
  Living Rev.\ Rel.\  {\bf 16}, 5 (2013).

\bibitem{AmelinoCamelia:1996pj}
  G.~Amelino-Camelia, J.~R.~Ellis, N.~E.~Mavromatos and D.~V.~Nanopoulos,
  Int.\ J.\ Mod.\ Phys.\ A {\bf 12}, 607 (1997).

\bibitem{AmelinoCamelia:1997gz}
  G.~Amelino-Camelia, J.~R.~Ellis, N.~E.~Mavromatos, D.~V.~Nanopoulos and S.~Sarkar,
  Nature {\bf 393}, 763 (1998).



\end{thebibliography}
\end{document}